\documentstyle[12pt]{article}
\title{
Comment on Vortex Mass and Quantum Tunneling of Vortices.}
\author{
 G.E. Volovik\\
 Low Temperature Laboratory, Helsinki University of Technology\\
 Otakaari 3A, 02150 Espoo, Finland\\
and\\
L.D. Landau Institute for Theoretical Physics, \\
Kosygin Str. 2, 117940 Moscow, Russia\\
}
\begin{document}
\maketitle
\begin{abstract}
{Vortex mass in Fermi superfluids and superconductors and its
influence on quantum tunneling of vortices are discussed. The
vortex mass is essentially enhanced due to the fermion zero modes
in the core of the vortex: the bound states of the
Bogoliubov qiasiparticles localized in the core. These bound states
form the normal component which is nonzero even in the low
temperature limit. In the collisionless regime $\omega_0\tau \gg
1$, the normal component trapped by the vortex is unbound from the
normal component in the bulk superfluid/superconductors and adds
to the inertial mass of the moving vortex. In the $d$-wave
superconductors, the vortex mass has an additional factor
$(B_{c2}/B)^{1/2}$ due to the gap nodes.}
\end{abstract}

The vortex mass is thought to be an important issue for the
problem of the quantum tunneling of  vortices. The latter
problem is popular now and many experiments are discussed in terms
of the macroscopic quantum tunneling of vortices in superfluids or
superconductors.  The firm experimental prove for the quantum
nucleation of vortices is still absent. On the other hand the
characteristic plateau in the temperature dependence of the
critical velocity, which is always ascribed to the quantum
nucleation, has been also observed in superfluid
$^3$He-B \cite{SingleVortexNucleation}. However the time of the
quantum nucleation of the vortex in  $^3$He-B is $10^{10^6}$, which
is extremely big in any units.  The vortices in $^3$He-B are created
in the process of the development of the classical instability of
the superflow at the pair-breaking velocity. The reason of the
plateau is that the characteristic physical quantities,  such as
the gap amplitude $\Delta$,  which determine the treshold of
instability   become  temperature independent at low $T$. The
intrinsic instability thus provides an alternative explanation of
the plateau observed in many different systems including
superfluid $^4$He.

In the vortex tunneling problem the inertial mass becomes
important only if its effect is comparable with the effect of the
Magnus force. That is why the magnitude of the inertial mass is of
main importance. It appears that in Fermi superfluids and
superconductors the mass of the vortex is essentially enhanced
compared to the vortex mass in superfluid
$^4$He, where it is determined by the compressibility. In Fermi
systems the fermions bound to the vortex core give the dominating
contribution, as was first found by Kopnin
\cite{Kopnin1978}. We discuss this effect in details and
relate it to the normal component trapped by the vortex. This
effect is even more enhanced in
$d$-wave superconductors where the vortex traps an essential part
of the bulk excitations due to the gap nodes.

{\it Volume law and area law for the quantum tunneling.}
In the earlier estimations of the vortex tunneling rate the mass of
vortex line was neglected \cite{Volovik1972,Sonin}. When the mass
is neglected the tunneling exponent $\exp{-S_{eff}}$ is determined
by the volume $V$ within the surface swept by the
classical trajectory of the vortex in the process of the quantum
tunneling:
 $$S_{eff}/\hbar= 2\pi {\cal N}~~,~~{\cal N}=nV ~.\eqno(1.1)$$
Here $n$ is the particle density; ${\cal N}$ is the number of
particles in the volume $V$. The volume law for the vortex action
follows from the general laws of the vortex dynamics governed by
the Magnus force \cite{Rasetti}.

In Ref.\cite{Volovik1972} the tunneling trajectory between the
ground state of the superfluid and the state with a vortex,
was generated  by irregularity (pinning center) on the container
wall in the presence of the superflow with the
asymptotic superfluid velocity $v_s$. For small $v_s$ the
tunneling exponent does not depend on the pinning center and
corresponds to the volume
 $$ V = {4\pi\over 3}
R_0^3~.\eqno(1.2)$$
Here $R_0$ is the radius of the nucleated
vortex ring:
$$R_0=(\kappa/2\pi v_s) \ln{R_0\over R_{core}}~ \eqno(1.3)$$
and  $R_{core}$ is the core size, which is of order
coherence length $\xi$.

The volume law for the tunneling exponent $S_{eff}$ was confirmed in
Ref.\cite{Sonin}, where   $S_{eff}$ was found as the overlapping
integral of the many-body wave function. This  $S_{eff}$ was
then minimized with respect to the velocity field in the vortex. The
extremal trajectory corresponds to the formation of the
intermediate  vortex state with the deformed velocity field around
the vortex loop. The resulting volume $V$ is logarithmically reduced
compared with the Eq.(1.2) for the direct formation of the
equilibrium vortex:
$$S_{eff}/\hbar= 2\pi nV ~~,~~
V ={27\over \pi  \ln{R_0\over R_{core}} }  R_0^3~.\eqno(1.4)$$
In this approach the volume law reflects the general property
of the macroscopic quantum tunneling: the tunneling exponent is
proportional to the number ${\cal N}$ of particles, effectively
participating in the tunneling. This was also found in other
systems \cite{LifshitzKagan,IordanskiiFin}.

When the problem of the vortex tunneling was revived due to the
experiments on the vortex creep in superconductors, the effect of
the vortex mass was discussed \cite{Blatter}.  If the mass term is
more important for the quantum tunneling than the Magnus force,
then  the  volume law  of Eq.(1.1) should transfer to the area law.
The quadratic dependence of $S_{eff}$ on $R_0$ (area law) was also
obtained using the field theory in \cite{Davis,Kao}, where the
vortex nucleation was considered as a process analogous to the
Scwinger production of the electron-positron pairs in electric
field.  The result for the semi-classical tunneling exponent is
$$S_{eff}=\int_0^{R_0} dR ~\sqrt{2 M (R)  E_{\rm
vortex }(R) } ~~,\eqno(1.5)$$
where
$E_{\rm vortex }(R)  \sim R \ln{R\over R_{core}}$ is the
energy of the vortex ring of radius $R$  and $M(R)$ is the
mass of the vortex loop. Since
$M(R)$ is also
$\propto R$ the tunneling rate is proportional to the area $R_0^2$
of the nucleated vortex ring.

This area law for the  action is typical for the dynamics
of string loops in systems without Magnus force, such as cosmic
strings (see \cite{LundRegge}), vortex
rings in charge-density-wave system
\cite{CDW}, in antiferromagnets, etc. The breaking of
the time inversion symmetry introduces the Magnus force even in
these systems (see Ref.\cite{SoninNikiforov} on vortices in
planar magnets and
Ref.\cite{DavisShellard} on spinning global strings), and
the volume law can be restored.

{\it  Hydrodynamic mass of the vortex.}
In the  hydrodynamic
theory  the mass of the vortex is nonzero due to compressibility of
the liquid which leads to the "relativistic" expression
\cite{Davis,Duan,Kao,WexlerThouless}
$$M_{\rm hydro}={E_{\rm vortex }\over s^2}~~, \eqno(2.1)$$
where $s$ is sound velocity. For Fermi superfluids $s$ is
of order the Fermi velocity
$v_F\sim p_F/m$ ($m$ is the mass of the electron or of
the $^3$He atom), and the estimation for the
hydrodynamic mass of the vortex loop of length
$L$ is
$$M_{\rm hydro} \sim L m k_F  \ln{L\over \xi}~~. \eqno(2.2)$$

However in this consideration the fermions in the vortex core
\cite{Caroli} are neglected.  They produce the effective mass
proportional to the core area
$ R_{core}^2 \sim \xi^2$
\cite{Kopnin1978,Kopnin1995,vanOtterlo1995,Stone}:
$$M_{\rm
bound~states}\sim L m k_F(k_F \xi)^2 ~~. \eqno(2.3)$$
Though it does not  contain the logarithmic divergence, it gives the
main contribution since the core radius
$\sim\xi$ in superfluid $^3$He-B and superconductors is large
compared with the interatomic space: $k_F \xi \gg 1$.
The mass of the vortex is essentially enhanced, so the
arguments, that the effect of the vortex mass  on the vortex
tunneling is negligible
\cite{Volovik1972,Stephen}, become shaky. That is why it is
worthwhile to consider the effect of core fermions more
thoroughly.

{ \it Bound states contribution to the vortex mass: Normal
component in the vortex core in collisionless regime.}
The core contribution to the vortex mass was obtained  by Kopnin
\cite{Kopnin1978} in a rigorous microscopic theory for the vortex
dynamics developed by Kopnin and Kravtsov \cite{Kopnin-Kravtsov}.
Here we associate it with the normal component trapped by the core
texture. At low
$T$ the core contribution to the vortex dynamics is
completely determined by the low-energy excitations in the vortex
core, which energy spectrum is
$E=-Q\omega_0(k_z)$ in the vortex frame \cite{Caroli}. Here $Q$ is
the angular momentum of fermions and
$\omega_0(k_z)$ is the interlevel spacing, which depends on the
linear momentum
$k_z=k_F\cos\theta$ along the vortex axis ($\omega_0 \sim
\Delta^2/E_F\ll \Delta$). If the temperature is large enough,
$\omega_0
\ll T \ll T_c$, this branch is characterized by the density of
states $N(0)=1/ \omega_0(k_z)$.

If the vortex moves with  velocity ${\bf v}_L$ with
respect to the superfluid component, the fermionic spectrum in the
vortex frame is Doppler shifted
$E=-Q\omega_0(k_z) - {\bf k}\cdot  {\bf v}_L$. In the collisionless
regime, $\omega_0\tau \gg 1$, the exchange between the fermions in
the vortex core and in the heat bath vanishes and the linear
momentum of the bound states fermions adds to the momentum of the
moving vortex.  The summation of fermionic momenta  in the
moving  vortex leads to the extra linear momentum of the
vortex  $\propto{\bf v}_L$ (see also
Eq.(5.7) in Ref.\cite{Stone}):
$${\bf P}=\sum {\bf k}\theta (-E)=M_{\rm bound~states}{\bf
v}_L ~~,\eqno(3.1)$$
$$M_{\rm bound~states}
=  L\int_{-k_F}^{k_F} {dk_z\over
4\pi}{k_\perp^2  \over
\omega_0(k_z)}  ~~,\eqno(3.2)$$
This is  an extra
vortex mass
which is by factor $(k_F\xi)^2$ larger than the hydrodynamical
mass of the vortex.

The Eq.(3.2) represents the dynamical mass of
the vortex in the low temperature limit and only in the clean (or
collisionless) regime,  when the exchange between the core fermions
and the heat bath is suppressed. Actually it was assumed that
$T_c\gg T\gg\omega_0\gg 1/\tau$. In this regime there is no spectral
flow between the bound fermions and the heat bath, as a
result during the vortex motion the momentum of core fermions is not
transferred to the heat bath and adds to the momentum of the
vortex, producing an extra inertia. In other words, this is  the
contribution of the normal component associated with the
vortex core, which in the collisionless regime is trapped by the
vortex and is transferred together with the vortex.

For vortices, which core size  $R_{core}\gg\xi$, this extra vortex
mass can be represented as the integral over the local density of
the normal component
$$M_{\rm bound~states}=\int d^3r ~\rho_n({\bf r},T=0)
~~.\eqno(3.3)
$$
This nonzero normal component at $T=0$ is
produced by the inhomogeneous order parameter, the texture. This
can be seen on the simplest example of continuous vortex
in $^3$He-A-phase, where the corresponding texture
is the field of the unit vector $\hat {\bf l}$ along the
orbital angular momentum of Cooper pairs. Let us choose the
texture in the form
$$\hat {\bf l}({\bf r})=\hat {\bf z}\cos\eta(r) + \hat {\bf
r}\sin\eta(r) ~~,\eqno(3.4)
$$
with $\hat {\bf l}(0)=-\hat {\bf z}$ and $\hat {\bf
l}(\infty)=\hat {\bf z}$. This texture represents the doubly
quantized continuous vortex in $^3$He-A (see Eq.(5.21) in Review
\cite{SalVol}), the latest experiments on such vortices are
discussed in \cite{Manninen}.

The $\hat {\bf l}$-texture leads to the normal
component tensor even at $T=0$
\cite{VolovikMineev1981} (see Eq.(5.24) in \cite{Volovik1990}):
$$(\rho_n)_{ij}({\bf r})={k_F^4 \over 2\pi^2 \Delta_A} \vert (\hat
{\bf l}\cdot \vec\nabla)\hat {\bf l}\vert ~\hat  l_i\hat l_j
~~,\eqno(3.5)
$$
where $\Delta_A$ is the gap amplitude in $^3$He-A. For
the texture in Eq.(3.4) one has
$\vert (\hat {\bf l}\cdot \vec\nabla)\hat {\bf l}\vert=\sin\eta~
\partial_r\eta$, so the normal component contribution to
the vortex mass should be
$$M_{\rm bound~states}\delta_{\perp ij}=\int d^3r
(\rho_n)_{ij}
=L{k_F^4 \over 2\pi \Delta_A}
\int_0^\infty dr ~r\sin^3\eta~
\partial_r\eta~~.\eqno(3.6)
$$

The Eq.(3.6) for the vortex mass in terms of the local normal
component coincides with the general Eq.(3.2) for the vortex
mass  in terms of
$\omega_0(k_z)$. The interelevel spacing for this  continuous vortex
was found by Kopnin \cite{Kopnin1995}
$$ \omega_0(k_z)={\Delta_A \over k_F
r(k_z)}~~,~~\cos\eta(r(k_z))={k_z\over k_F}.~~\eqno(3.7)
$$
Here $r(k_z)$ is the radius where the energy of the fermion
$$E(r,\vec k)=\sqrt{v_F^2(k-k_F)^2 +\Delta_A^2(\hat {\bf
l}(r)\times\hat {\bf k})^2}$$  is zero at given
$k_z$.   The Eq.(3.2) gives  \cite{Kopnin1995}
$$M_{\rm bound~states} =  L\int_{-k_F}^{k_F}
{dk_z\over 4\pi}{k_\perp^2  \over
\omega_0(k_z)}={k_F\over 4\pi \Delta_A }\int_{-k_F}^{k_F}
dk_z (k_F^2 -k_z^2) r(k_z).$$
After inverting the function $r(k_z)$ in Eq.(3.7) into
$k_z(r)=k_F\cos\eta(r)$ one obtains the Eq.(3.6).

{\it Vortex mass from the kinetic equation.}
The above results for the vortex mass can be proved using the
kinetic equation for the  fermions bound to the core
\cite{Kopnin1978,Kopnin1995,vanOtterlo1995}. The inertial
term in the force balance for the vortex is obtained by
substitution
${1\over
\tau} $  by  ${1\over
\tau} -i\omega$ in the equation for the longitudinal (dissipative
friction) force acting on the vortex line, where
$\omega$ is external frequency identified with the frequency of the
oscillations of the vortex line. In the
temperature region
$\omega_0\ll T
\ll T_c$ one has \cite{Kopnin1995}
$${\bf F}_{\rm long}=- {\bf v}_L{k_F^3\over
4\pi}L\int  d\cos\theta ~ \sin^2\theta ~\left({1\over
\tau} -i\omega\right) {\omega_0\over
\omega_0^2 +\left({1\over
\tau} -i\omega\right)^2}
  ~.\eqno (3.8)$$

In the limit case $\omega_0\gg \omega\gg 1/\tau$ one obtains
${\bf F}_{\rm long}=  i\omega {\bf v}_L M_{\rm bound~states}$
with the vortex mass
$$M_{\rm bound~states}= { 3\pi \over 4} L C_0\int  d\cos\theta ~
\sin^2\theta ~   {1
\over
\omega_0(\theta)}~~,\eqno (3.9)
$$
where $C_0=k_F^3/3\pi^2$ is close to
the particle density $n$. This corresponds to Eq.(3.2).

In the high frequency limit $\omega\gg \omega_0\gg 1/\tau$ the
Eq.(3.8) leads to the "dielectric" behavior with the "pinning
potential"
$$U= {1\over 2} \alpha {\bf r}_L^2~~,~~\alpha= {k_F^3\over
4\pi}\int  d\cos\theta ~ \sin^2\theta ~
\omega_0(\theta)~~.\eqno (3.10)
$$

{\it Discussion.}
The vortex inertia is essentially enhanced due to the
fermion zero modes in the vortex core. This fermionic contribution
to the vortex mass appears when the characteristic frequency is
small compared to the interlevel distance $\omega\ll
\omega_0$. The characteristic frequency of the tunneling process
can satisfy this condition, since $\omega\sim   \sqrt{E_{\rm
vortex }(R_0)/M(R_0)R_0^2} \sim \omega_0\xi/R_0 <\omega_0$.
If $\omega >\omega_0$ the more general contribution of the core
fermions, the Eq.(3.8), is to be applied. But even in this case the
effect of the fermions is always larger then the
contribution of the hydrodynamic mass in Eq.(2.1).  This is because
the  frequency $\omega$ of the vortex motion cannot exceed the
magnitude of the gap $\Delta$, otherwise the simple approach to the
vortex dynamics is not valid. This means that the hydrodynamic mass
in Eq.(2.1) never enters the tunneling rate in Fermi superfluids and
superconductors.

On the other hand because of the limited frequency the effect of
the inertial mass on the vortex tunneling is still small compared
to the effect of the Magnus force.  Since
$\omega \ll\omega_0$
the kinetic term
$M\dot {\bf v}_L=-i\omega M {\bf v}_L \sim \hbar  n L
(\omega/\omega_0)$ is always smaller than the Magnus force $\pi
\hbar  n L \hat z
\times {\bf v}_L$. That is why the volume law for the tunneling
exponent in Eq.(1.1) is still dominating.

Situation can change in the regime $\omega_0\tau
\ll 1$, where the Magnus force is suppressed by the
spectral flow of fermions: $\pi
\hbar  n L \hat z
\times {\bf v}_L \rightarrow  \pi
\hbar  (n-C_0) L \hat z
\times {\bf v}_L$
\cite{Q-modes-Index,KopninVolovik,vanOtterlo1995,Stone}.

The vortex mass can be also important in the
$d$-wave superconductor, where the effect of the
fermions on the vortex is
more pronounced due to gap nodes \cite{d-waveVortex}.
In these superconductors with highly anisotropic gap the
interlevel spacing depends on the azimuthal angle $\alpha$
between the momentum ${\bf k}$ in the $a-b$ plane and the direction
of the gap nodes \cite{d-waveVortex}
$$ \omega_0(\alpha)\approx  \alpha  ^2~{ \Delta_0^2\over
E_F}\ln {1\over
\vert  \alpha  \vert}~~,\eqno(4.1)$$
where $\Delta_0$ is the gap amplitude.
The vortex mass in Eq.(3.2) is:
$$M\delta_{\perp ij} =  L\int_{-k_F}^{k_F}
{dk_z\over 2\pi}\int_{0}^{2\pi}
{d\alpha\over 2\pi}k_{\perp i} k_{\perp j} {1\over
\omega_0(k_z,\alpha)}  ~~.\eqno(4.2)
$$
With Eq.(4.1) for $ \omega_0(\alpha)$ the integral over $\alpha$
diverges near the gap nodes. The cut-off
$\alpha_{min}\sim \xi/R_v$, where $R_v\sim \xi \sqrt{B_{c2}/B}$ is
the intervortex distance, gives the $\sqrt{B_{c2}/ B}$ enhancement
of the vortex mass:
$$ M\sim  m k_F^3 \xi^2 L  \sqrt{B_{c2}\over B} ~~.\eqno(4.3)
$$
This equation holds if
$1\gg  {B/B_{c2}}  \gg T^2/T_c^2$ and $  {B/B_{c2}} \gg
E_F/\tau \Delta_0^2$.

I thank N.B. Kopnin for illuminating discussions.
This work was supported through the ROTA co-operation plan of the
Finnish  Academy and the Russian Academy of Sciences and by the
Russian Foundation for Fundamental Sciences,  Grant No.
96-02-16072.

\end{document}